\documentclass[12pt]{article}
\title{ Remark on the effective range approach to analyzing  the
$f_0-a_0$ mesons
\\}
\author{B.Kerbikov,\\
 State Research
Center\\Institute of Theoretical and Experimental Physics, \\
Moscow, Russia}
 \date{}
  \newcommand{\be}{\begin{equation}}
\newcommand{\ee}{\end{equation}}
\def\la{\mathrel{\mathpalette\fun
<}} 
\def\fun#1#2{\lower3.6pt\vbox{\baselineskip0pt\lineskip.9pt
\ialign{$\mathsurround=0pt#1\hfil ##\hfil$\crcr#2\crcr\sim\crcr}}}

\newcommand{\ver}{\mbox{\boldmath${\rm r}$}}

\newcommand{\vep}{\mbox{\boldmath${\rm p}$}}

\newcommand{\lan}{\langle}
\newcommand{\ran}{\rangle}

\begin{document}
\maketitle

\begin{abstract}
An over simplified analysis of issues related to a resonance close
to threshold may  lead to misleading results. We clarify some
subtle points, in particular the relation between the Breit-Wigner
and effective range approaches to the $f_0-a_0$ mesons.
\end{abstract}

The $f_0$ and $a_0$ scalar mesons present a well-known puzzle for
which several interesting, albeit controvertial proposals have
been made ranging from quark-antiquark makeup to four-quark and to
$K\bar K$ molecular structure. The question of how to distinguish
between different hypotheses for the $f_0-a_0$ nature has been
discussed in a great number of papers and it is not our goal to
review all these suggestions. Long ago it was proposed to use the
sign and the value of the $K\bar K$ effective range parameter as a
signature of the  $f_0-a_0$ makeup \cite{1}. Namely, it was
suggested that the large and negative value of $r_e$ corresponds
to the case of a large admixture of the bare quark state while a
small (positive or negative) $r_e$ is a sign that the resonance
contains a large mesonic component \cite{1}. The  question  of
whether it is possible to judge  the nature of the $f_0-a_0$
mesons having at hand the values of the low-energy $K\bar K$
parameters has been revisited in a recent arXive submission
hep-ph/0308129 \cite{2}. In our opinion some remarks on this,
presented below,  significantly   qualify the answer to question
regarding  the distinguishability of the $f_0-a_0$ makeup on the
basis of their phenomenological parameters. In this note we do not
 discuss the question of  to what extent the  value and the sign of
the effective range are   sound signatures of  resonance makeup.
What will be shown is that  scrutiny of the effective range
calculations presented in \cite{2} results in serious doubts about
their reliability.

Equation (20 ) of \cite{2} for the $K\bar K$ effective range reads
\be r_e=-\frac{4}{m\bar g_{K\bar K}}.\label{1}\ee

Here $m=(m_{K^+}m_{K^0})/2, \bar g_{K\bar K} = g^2_{K\bar K}/8\pi
M^2$, where $M$ is the mass of the resonance $(a_0$ or $f_0$), and
$g^2_{K\bar K}$ is the standard coupling constant connected to the
resonance width via $\Gamma_{K\bar K} = g^2_{K\bar K}(8\pi
M^2)^{-1}k$, where $k$ is the c.m. momentum. For definiteness in
what follows we consider only the $f_0$- case.

The value of the coupling constant $\bar g_{KK}$  suffers a large
ambiguity, due to the  variety of  experimental data and the lack
of consensus in its definition. Hence the results for the
effective range presented in Table 2 of \cite{2} and calculated
according to (\ref{1})  span  from  -0.56 fm to -1.22 fm. One may
add another line to the Table 2 of \cite{2} using the value of
$\bar g_{K\bar K}$ presented in \cite{3} (set $D$ of Table 1   in
\cite{3}); then equation (\ref{1}) yields $r_e =-4.79 $ fm.
Following the arguments of \cite{1,2} one should conclude that the
value of $r_e=-0.56$ fm corresponds to a large admixture of the
$K\bar K$ component while $r_e=-4.79$ fm is  evidence for  a
dominant quark nature  of  the  $f_0$-- meson. We do not want to
discuss this obvious contradiction. Our next step is to replace
equation (\ref{1}) by the one  which  correctly takes into account
the interplay of a quark state and a hadronic threshold. We shall
show that, with $\bar g_{K\bar K}$ constants  the same as in
\cite{2},
 the value $r_e=-0.56$ fm is replaced by $r_e\simeq
+0.8$ fm, while instead of $r_e=-4.79$ fm one gets $r_e\simeq
-3.5$ fm.

The problem may be approached in different ways. In particular, a
correct treatment is presented in \cite{4}, where one  finds a
simple model for the interplay of quark and hadronic channels in
the $f_0$-meson. Here we follow an alternative approach based on
the formalism developed  in \cite{5}-\cite{8}.   Consider a
three-channel system with two hadronic channels denoted by $\pi$
(the $\pi\pi$) and $K$ (the $K\bar K$) and a quark channel denoted
by $q$ ($q\bar q$ or $q^2\bar q^2$). As we shall see in a moment,
the $\pi\pi$ channel plays no dynamical role in the model and
therefore  relativistic treatment which   is  formally needed for
the $\pi\pi$ system would  not change anything. The quark channel
is parametrized by the position of the bare level $E_n$ and the
communication potentials $V_{qi},$ $i=\pi,K$. We assume that there
is no residual potential interaction between the two hadrons
($\pi\pi$ or $K\bar K$). The generalization of this model to the
case of several levels in the  quark channel and the  hadronic
interaction may be found in \cite{7,8}. The exact equation for the
$S$-matrix in the channel $K$ reads \cite{7,8}
\be
S_K(E)=\frac{E-E_n+\lan q|V_{q\pi} G^{(+)}_\pi V_{\pi q}|q\ran+
\lan q|V_{qK}G_K^{(-)}V_{Kq}|q\ran}{E-E_n+\lan q|V_{q\pi}
G^{(+)}_\pi V_{\pi q}|q\ran+ \lan
q|V_{qK}G_K^{(+)}V_{Kq}|q\ran},\label{2}\ee where the indices
$(+)$ and  $(-)$ in (\ref{2}) correspond to the choice of the
boundary conditions $\pm i0$ in the Green's functions of the
corresponding channels.  Since  we are interested in the phenomena
taking place  in  the energy interval of the order of a few tens
of MeV around the $K\bar K$ threshold, we may neglect the energy
dependence of the $\pi\pi$ matrix element and parametrize it in a
standard way
\be
\lan q|V_{q\pi}G_\pi^{(+)}V_{\pi q}|q\ran=-\varepsilon_{\pi\pi}
+\frac{i}{2} \Gamma_{\pi\pi},\label{3}\ee where
$\varepsilon_{\pi\pi}$ is  the hadronic shift of the level and
$\Gamma_{\pi\pi}$ is the width of  $ f_0$-meson  into $\pi\pi$. We
shall include the constant shift $\varepsilon_{\pi\pi}$  into the
energy of the level $E_n$. The energy dependence of a similar
matrix element for the $K\bar K$ channel is crucial. Again as in
(\ref{3}) we represent the matrix element for the $K\bar K$
channel as a sum of  Hermitian and anti-Hermitian parts depending
upon the $K\bar K$ c.m. momentum $k$
\be
\lan q|V_{qK}G_K^{(\pm)}V_{Kq}|q\ran=- \varepsilon_{K\bar K} (k)
\pm \frac{i}{2} \Gamma_{K\bar K} (k).\label{4}\ee

Equations (\ref{2})-(\ref{4}) yield the following $T$-matrix for
the $K\bar K$ channel
\be
T_K=(2\pi^2mk)^{-1} \frac{\Gamma_{K\bar K}(k)/2}{E-E_n+\frac{i}{2}
\Gamma_{\pi\pi} -\varepsilon_{K\bar K} (k) +\frac{i}{2}
\Gamma_{K\bar K}(k)}.\label{5}\ee

At this point it is tempting to parametrize $\Gamma_{K\bar K}(k)$
as $\Gamma_{K\bar K} (k) =\bar g_{K\bar K} k$ and to include
$\varepsilon_{K\bar K} (k)$ into $E_n$ (as it was previously done
with $\varepsilon_{\pi\pi}$). Then one arrives at Eq. (17) of
\cite{2}, namely
\be
f_K =-\frac{\bar g_{K\bar K}/2}{\frac{k^2}{m} -E_n + \frac{i}{2}
\Gamma_{\pi\pi} +\frac{i}{2} \bar g_{K\bar K} k},\label{6}\ee
which immediately leads to the expression (\ref{1}) for the
effective range $r_e$.

What was done wrongfully in passing from (\ref{5}) to (\ref{6})?
The answer is clear: the term proportional to $k^2$ stemming from
$\varepsilon_{K\bar K} (k)$ was   omitted. This omission is
mentioned in  \cite{2} without  further discussion. When included,
this term adds to the expression (\ref{1}) the contribution which
is  of the same order as (\ref{1}) but of the  opposite sign. To
see this let us return to Eq. (\ref{4}) and write the matrix
element explicitly $$-\varepsilon_{K\bar K} (k) +\frac{i}{2}
\Gamma_{K\bar K} (k) = \int d\vep d\vep'\lan q|V_{qK}|\vep \ran
\frac{\delta(\vep-\vep')}{\frac{p^2}{m}-E(k)-i0} \lan
\vep'|V_{Kq}|q\ran= $$
\be
= 4\pi m \int d p p^2\frac{|\lan q|V_{qK}| \vep\ran
|^2}{p^2-k^2-i0}.\label{7} \ee The result (\ref{1}) of \cite{2}
for $r_e$ corresponds to the  assumption that the principal value
of the  integral (\ref{7}) does not depend on $k^2$. Then  the
constant shift $\varepsilon_{K\bar K}$ may be absorbed into $E_n$
as it was done with $\varepsilon_{\pi\pi}.$

Let us demonstrate that this assumption breaks down. To make
(\ref{7}) easily tractable consider the most widely used model for
the communication potential $V(r)$, namely
\be
\lan r |V_{Kq}|q\ran =\gamma^{1/2}
\frac{\delta(r-b)}{b\sqrt{4\pi}},\label{8}\ee where $\gamma$ is a
constant with the dimension of mass, and $b$ is the range at which
the transitions between the channels effectively occur. The
formfactor in momentum space corresponding to the transition
potential (\ref{8}) reads
\be
\lan q|V_{qK}|k\ran =\int d\ver \lan q|V_{qK}|\ver \ran \lan \ver
|k\ran =\frac{\gamma^{1/2}}{\pi\sqrt{2}} \frac{\sin
kb}{k}.\label{9}\ee Therefore we are dealing with a smooth
transition formfactor in momentum space with the range $k\sim
\pi/2b$, i.e. $b\simeq\pi/2\beta$ in the notations of  Ref.
\cite{2}.
 Substituting (\ref{9}) into the integral (\ref{7}), one gets
 $$
 -\varepsilon_{K\bar K}(k) + \frac{i}{2} \Gamma_{K\bar K}(k) =
 \gamma \frac{m}{k} e^{ikb} \sin  kb\simeq
 $$
 \be
 \simeq \gamma mb (1-\frac23 k^2 b^2) + im\gamma b^2
 k.\label{10}\ee

 If, following \cite{2}, we use the parametrization $\Gamma_{K\bar
 K}(k)= \bar g_{K\bar K} k$, we have to identify
 \be
 \bar g_{K\bar K} = 2\gamma m b^2.\label{11}\ee
 From (\ref{5}) and (\ref{10}) the effective range is easily
 calculated to be
 \be
 r'_e =-\frac{4}{m\bar g_{K\bar K}} + \frac43 b = r_e +\frac43
 b,\label{12}\ee
 where $r_e$ is given by Eq. (\ref{1}) which is identical to Eq.
 (20) of \cite{2}.
 The authors of Ref.\cite{2} point out that with Eq. (\ref{1}) it
 is impossible to reproduce the  deuteron-like situation with
 positive effective range. Eq.(\ref{12}) is free of this deficiency.

 Now we return
to Eq.(\ref{8}) and discuss the physics behind it. The formal
multichannel scattering theory with the communication potential
(\ref{8}) was developed in Ref. \cite{6}. Less rigorous approach
was followed in \cite{7} and in \cite{8}. Historically the use of
 the boundary condition (\ref{8}) probably goes back to a seminal
 paper  by  C.Bloch \cite{9}. Numerous calculations of different
 quark-hadron systems based on (\ref{8}) were performed - see e.g.
 \cite{10, 11, 12}. A very important observation was done in
 \cite{13}, namely that the model with the $\delta$ -function
 transition potential (\ref{8}) is equivalent to the Jaffe- Low
 P-matrix \cite{14}. Therefore  we may use the well-known
 $P$-matrix recipes to estimate the range $b$ and the coupling
 constant $\gamma$. For   meson-meson system this yields
 \cite{4,14,15,16}
 \be
 b\simeq 1.4 R, ~~R\simeq 5 M^{1/3} {\rm GeV}^{-1}, \label{13}\ee
 where $M$ is the mass of the bare quark state,  $M\simeq 1$ GeV
 for   $f_0$ -meson, i.e. $R\simeq  1$ fm. The absolute lower
 bound  on $b$ is $b_{min} = 0.4 R\simeq 0.4$ fm \cite{4}. As it
 was mentioned after  (\ref{9}) the corresponding range of the
 transition formfactor in momentum space is $k\equiv \beta \simeq
 \pi/2b_{min}\simeq 800$ MeV. Obviously the nonrelativistic
 approach used in \cite{2} and in the present work completely
 breaks at such values of the relative $K\bar K$ momenta. Therefore a
 conservative estimate of $b$ is $b\simeq 1 fm \simeq 5 $
 GeV$^{-1}$. According to (\ref{12}) it means that the values of
 the effective range are 1.3 fm larger than the results presented
 in Table 2 of \cite{2}. The values of $r_e$ listed in this Table
 range from -0.56 fm to -1.22 fm, and hence we are dealing with
 $\ran 100$\% "correction".

 The  connection  between the formfactor (\ref{8}) and $P$-matrix
 allows to estimate the coupling constant $\gamma$. It is related
 to the residue $\lambda_{K\bar K}$ of the $P$-matrix with respect
 to the $K\bar K$ channel via $\Gamma=\lambda_{K\bar K}/m$
 \cite{17}. This residue is known  very approximately,
 $\lambda_{K\bar K} \simeq 0.02$ GeV$^2$ \cite{16}, so that $\gamma\simeq 0.04 $ GeV. According to (\ref{11})
 this  corresponds to $\bar g_{K\bar K}\simeq 1$,
   but this result relies
    on the rather uncertain  value of $\lambda_{K\bar K}$.
Once more we see that the determination of the coupling constant
$g^2_{K\bar K}(f_0)$ is  a problem still waiting for its solution.

 The situation with $P$-matrix parameters is much more clear in
 the $NN$ sector. As an example consider the set of parameters for
 $^3S_1$ state from Ref. \cite{18}, namely $\tau=2 m^2_N\gamma
 =0.4 $ GeV$^3$, $b=7.16$ GeV$^{-1}$. According to (\ref{11})
 and (\ref{1}) one gets $r_e=-4/\tau b^2=-0.039$ fm, while the
 correct Eq. (\ref{12}) yields $r'_e=1.87$ fm, which is close to
 the standard result $r_e=1.75$  fm  \cite{19}.

      We  also note that from the general expression
      (\ref{2}) for the $S$- matrix  one can  calculate the
       relative weights of  quark and hadron components in the physical $f_0$ -meson.
  The corresponding equation may be found in  Ref.\cite{8} (Eq. (19)) and in Ref. [20].

One may  ask a question to what extent are the effective range
calculations presented above model dependent. In particular,
whether the second  term in Eq. (\ref{12}) is really important, or
it contributes a minor correction in line with the statement of
Ref. \cite{2}. To clear out possible doubts let us turn to the
model independent calculation of
 the one-loop scalar propagator, or 1PI two-point function
 [3,21,22]. Using the standard  dimensional regularization
 of the loop diagram we find the following expression for the
 finite part of the inverse propagator
 \be
 D(s) = s-M^2 +\Sigma (s),\label{14}
 \ee
 \be
 \Sigma(s) =\frac{g^2_{K\bar K}}{16\pi} \left\{ i\rho +\frac{1}{\pi}
 \left [ 2-\rho\ln \frac{1+\rho}{1-\rho}\right]
 \right\},\label{15}\ee
where we have returned to the  dimensionfull coupling  constant
$g^2_{K\bar K}$, and where $\rho=2k/\sqrt{s}\simeq k/m$. The
contribution of the pion loop has been omitted. The
nonrelativistic reduction of (\ref{14}) -(\ref{15}) reads \be D(E)
\simeq -2 m \bar g_{K\bar K}\left \{\left( \frac{2}{\bar g_{K\bar
K}}E_n-\frac{2m}{\pi}\right) +\frac12 \left( -\frac{4}{\bar g m}
+\frac{4}{\pi m} \right) k^2 - ik\right\}.\label{16}\ee

From (\ref{16}) one obtains
\be
r'_e=- \frac{4}{\bar g_{K\bar K} m} \left( 1 - \frac{\bar g_{K\bar
K}}{\pi}\right) = r_e \left(1- \frac{\bar g_{K\bar
K}}{\pi}\right).\label{17}\ee

We have recovered the same structure of $r'_e$ as the one which is
given by the model (\ref{8}). The  additional term $\bar g_{K\bar
K}/\pi$ varies from 0.4 to 0.9 for the values of $\bar g_{K\bar
K}$ from the Table 2 of Ref.\cite{2}. Comparing (\ref{12}) and
(\ref{17}) we conclude that they are equivalent provided $b=3/\pi
m$ which corresponds to $b_{min}$ (see the text after (\ref{12})).
This is not surprising since the only scale parameter with the
dimension of length in the loop diagram is $1/m$. Physically more
sensible estimate is $b\simeq M^{1/3}$ GeV$^{-1} \simeq 1$ fm
\cite{4,14,15,16}.

Finally we wish to  note that Coulomb effects in $K^+ K^-$ system
have been neglected. They become really  important for $k\la
2\pi/a_B = \pi\alpha m$, where $a_B$ is the Bohr radius of the
$K^+K^-$ atom and $\alpha=1/137$. Therefore the expression
(\ref{2}) for the $S$-matrix has to be modified in the energy
interval $ E\la 0.3$ MeV. The correct form of the $f_0$ -meson
propagator with Coulomb interaction included  was  derived in Ref.
\cite{22}, and in Ref \cite{17} the interplay of the $f_0$ -meson
and $K^+K^-$ atom  was described.

We have  shown  that the expression (\ref{1}) for the effective
range undergoes a substantial change due to the term proportional
to $k^2$ arising from $\varepsilon_{K\bar K}(k)$. Next task would
be to reconsider in  a similar  way other quantities depending on
the $T$-matrix (\ref{5}), e.g. the spectral densities of the
$f_0/a_0$ mesons \cite{2}, or the positions of the poles \cite{4}.

In conclusion, we may repeat the general statement that the
problem of the $f_0$ - meson makeup is far from being resolved. It
is possible that more information may be obtained using quantum
mechanical approach. In particular the effective range parameter
may be an important quantity. The accurate evaluation of this
parameter  has been given  in the present note.

We would like to thank Yu.S.Kalashnikova and A.E.Kudryavtsev for
useful discussions. Interesting remarks by D.Bugg, F.Kleefeld and
P.Landshoff are gratefully acknowledged. Financial support from
the grant Ssc-1774-2003 is gratefully acknowledged.


\begin{thebibliography}{99}
\bibitem{1} N.A.Tornqvist, Phys. Rev. {\bf D51}, 5312 (1995).

\bibitem{2} V.Baru, J.Haidenbauer, C.Hanhart, Yu.Kalashnikova,
A.Kudryavtsev, Phys. Lett. {\bf B586}, 53 (2004).

\bibitem{3}R.Escribano et.al. Eur. Phys. J. {\bf C28}, 107 (2003).

\bibitem{4} S.V.Bashinsky and R.L.Jaffe, Nucl. Phys. {\bf A625}, 167 (1997).

\bibitem{5}
R.F.Dashen, J.B.Healy and I.J.Muzinich, Phys. Rev. {\bf D14} ,
2773 (1976).

\bibitem{6}
R.F.Dashen, J.B.Healy and I.J.Muzinich,  Ann. Phys.  {\bf 102}, 1
 (1976).

\bibitem{7} B.O.Kerbikov, Theor. and Math. Phys.., {\bf 65}, 1225
(1985).

\bibitem{8} B.O.Kerbikov, Quantum Mechanics of a System with
Confinement, Preprint ITEP-58, 1985.

\bibitem{9} C.Bloch, Nucl. Phys. {\bf 4}, 503 (1957).

\bibitem{10} C.Dullemond and E.van Beveren, Ann. Phys. {\bf 105},
318 (1977).

\bibitem{11} E.van Beveren, C.Dullemond, T.A.Rijken, Z.Phys.  {\bf
C19}, 275 (1983).

\bibitem{12} B.O.Kerbikov, Sov. J. Nucl. Phys. {\bf 41}, 461
(1985).

\bibitem{13} Yu.A.Simonov, Phys. Lett., {\bf B107}, 1 (1981);
Nucl. Phys., {\bf A463}, 231 c (1987).

\bibitem{14} R.L.Jaffe and F.E.Low, Phys. Rev., {\bf D19}, 2105
(1970).

\bibitem{15} A.K.A.Maciel and J.E.Paton, Nucl. Phys. {\bf B181},
277 (1981).

\bibitem{16} R.P.Bickerstaff, Phil. Trans. R.Soc. Lond., {\bf
A309}, 611 (1983).

\bibitem{17} B.Kerbikov, Z.Phys. {\bf A353}, 113 (1995).

\bibitem{18} Yu.A.Simonov, Preprint ITEP-143, 1981.

\bibitem{19} E.L.Lomon and R.Wilson, Phys. Rev. {\bf C9}, 1329
(1974).
\bibitem {20a} N.A.Tornqvist, Z.Phys. {\bf C68}, 647 (1995).

\bibitem{20} N.N.Achasov and V.V.Gubin, Phys. Rev., {\bf D56},
4084 (1997).

\bibitem{21} T.Bhattacharya and S.Willenbrock, Phys. Rev. {\bf
D47}, 4022 (1993).

\bibitem{22} S.V.Bashinsky and B.O.Kerbikov, Phys. Atom. Nucl.
{\bf 59}, 1979 (1996).

\end{thebibliography}
\end{document}